\documentclass[aps,prd,nofootinbib,showkeys,floatfix,twocolumn,bibnotes,preprintnumbers]{revtex4-1}
\usepackage{graphicx,amsmath,amssymb,hyperref,natbib,slashed}
\usepackage[dvipsnames]{xcolor}

\usepackage{xspace}
\usepackage{bbm} 
\usepackage{lipsum}
\usepackage{subfigure}

\usepackage[utf8]{inputenc}

\def\sfrac#1#2{{\textstyle{#1\over #2}}}
\newcommand{\be}{\begin{equation}}
\newcommand{\ee}{\end{equation}}
\newcommand{\ba}{\begin{array}}
\newcommand{\ea}{\end{array}}
\newcommand{\bea}{\begin{eqnarray}}
\newcommand{\eea}{\end{eqnarray}}
\newcommand{\sss}{\scriptscriptstyle}

\newcommand{\nn}{\nonumber}

\newcommand{\ds}{{\sf DarkSUSY}}

\begin{document}

\title{Baryogenesis from neutron-dark matter oscillations}
\author{Torsten Bringmann}
\email{torsten.bringmann@fys.uio.no}
\affiliation{Department of Physics, University of Oslo, Box 1048, NO-0371 Oslo, Norway}
\author{James M.\ Cline}
\email{jcline@physics.mcgill.ca}
\affiliation{McGill University, Department of Physics, 3600 University St.,
Montr\'eal, QC H3A2T8 Canada}
\author{Jonathan M.\ Cornell}
\email{jonathan.cornell@uc.edu}
\affiliation{McGill University, Department of Physics, 3600 University St.,
Montr\'eal, QC H3A2T8 Canada}
\affiliation{Department of Physics, University of Cincinnati, Cincinnati, Ohio 45221, USA}
\begin{abstract}

It was recently suggested  that dark matter consists of $\sim$GeV
particles that carry  baryon number and mix with the neutron. We
demonstrate that this could allow for  resonant dark matter-neutron
oscillations in the early universe, at finite temperature,  leading to
low-scale baryogenesis starting from a primordial dark matter
asymmetry.  In this scenario, the asymmetry transfer happens around 30
MeV, just before big bang  nucleosynthesis. We illustrate the idea
using a model with a dark $U(1)'$ gauge interaction,  which has
recently been suggested as a way of addressing the neutron lifetime
anomaly.  The asymmetric dark matter component of this model is
both strongly self-interacting and  leads to a suppression of matter
density perturbations at small scales, allowing to  mitigate the
small-scale problems of cold dark matter cosmology. Future CMB
experiments will be able to consistently probe, or firmly exclude,
this scenario.

\end{abstract}
\maketitle


\section{Introduction}

One of the curious coincidences of the $\Lambda$CDM cosmological 
model is the rough similarity of the contributions from baryons and
dark matter (DM) to the energy density, $\Omega_b \simeq 0.049$ versus
$\Omega_{\rm CDM} \simeq 0.26$ \cite{Aghanim:2018eyx}, given that in 
typical scenarios of the early
universe they have a completely different origin.  An appealing
feature of many models of asymmetric DM is that this
coincidence could be due to DM having a mass at the GeV scale, like
baryons, and a shared mechanism for the generation of the two
asymmetries.  This could be achieved through either a common 
``cogenesis'' event that creates the required asymmetries in the dark and 
visible sector, or through an efficient sharing of asymmetries 
created in independent ways, though (dark) sphalerons, new renomalizable interactions, or higher-dimensional 
effective operators (for a review see \cite{Zurek:2013wia}). 
In particular, the baryon asymmetry of the universe (BAU) could
be fully explained by the ``darkogenesis'' of an asymmetry 
in the dark sector that is subsequently transferred to the 
standard model (SM) sector. Ref.~\cite{Shelton:2010ta}
analyzes this scenario in terms of the lowest-dimension,
gauge-invariant effective operators that would allow such a
redistribution of a primordial asymmetry starting in the dark sector.
For example, the operator $\chi udd/\Lambda^2$ would induce roughly
equal asymmetries between DM particles $\chi$ and quarks if it was 
in equilibrium at high temperatures.\footnote{This 
example requires $m_\chi < m_n$ since it would 
otherwise allow the DM to decay into quarks.}

In this work, we consider a novel situation where the 
$\chi udd/\Lambda^2$ coupling only arises below the QCD phase
transition.  Then it is replaced by the 
unique relevant operator that could
connect Dirac DM to the SM, namely mass-mixing with the
neutron,
\be
   {\cal L}_{\rm mix} = - \delta m\ \bar n \chi + {\rm h.c.}
\label{Lmix}
\ee
Our purpose is to show that it can produce
 the baryon asymmetry at low temperatures $\sim 30$\,MeV,\footnote{We note that the residual symmetric baryon component
is negligible by this time; see 
{\it e.g.}~Fig.~5 of Ref.~\cite{Aitken:2017wie}.}
starting from a $\chi$ asymmetry (whose darkogenesis origin we do not try to
specify here).  This comes about by oscillations between $\chi$ and
$n$, in analogy to neutrino oscillations, that are resonantly enhanced
by finite-temperature effects. Oscillations of neutrons to a mirror sector partner have been 
studied extensively in \cite{Berezhiani:2005hv, Berezhiani:2006je,Berezhiani:2008bc,Berezhiani:2018eds}, 
however, to our knowledge these oscillations have never been considered 
as a means for baryogenesis.  This adds a new option to the two broad classes 
of asymmetry transfer 
mechanisms discussed so far,
namely (dark) sphalerons (\textit{e.g.} \cite{Barr:1990ca, Barr:1991qn, Kaplan:1991ah, Gudnason:2006yj, Shelton:2010ta, Buckley:2010ui, Blennow:2010qp, Dutta:2010va, Perez:2013tea}) and  renormalizable interactions or higher-dimensional operators (\textit{e.g.} \cite{Foot:2003jt, Hooper:2004dc, Kaplan:2009ag, Shelton:2010ta, Haba:2010bm, Buckley:2010ui, Perez:2013nra, Servant:2013uwa, Bernal:2016gfn}). 
Similiar ideas for producing the 
baryon asymmetry at low temperatures via oscillations of SM baryons have been
recently suggested in Refs.~\cite{Aitken:2017wie,Elor:2018twp}.

Recently motivation for the operator in Eq.~(\ref{Lmix}) came from a quite
different direction.  It was suggested \cite{Fornal:2018eol} that mixing between the neutron
and DM could resolve a long-standing discrepancy between determinations of
the neutron lifetime from decay-in-flight ($p$ appearance) versus bottle 
($n$ disappearance) measurements,
by having a small dark decay channel \cite{Wietfeldt:2011suo}.\footnote{%
It has also been proposed that neutron-mirror neutron oscillations could explain 
this puzzle \cite{Berezhiani:2018eds}.
} 
The proposed decays $n\to
\chi\gamma$ and $n\to\chi e^+ e^-$ were quickly ruled out by
experimental searches \cite{Tang:2018eln,Sun:2018yaw}, suggesting a
completely hidden channel, like $n\to\chi\gamma'$ where $\gamma'$
is a dark photon.  Even this model is ruled out by neutron star 
properties \cite{McKeen:2018xwc,Baym:2018ljz,Motta:2018rxp,Gandolfi:2011xu}, 
unless repulsive $\chi$ self-interactions are strong enough,
requiring 
\be
	{m_{\gamma'}\over g'} \lesssim (45-60)\,{\rm MeV}
\label{NS}
\ee
to be satisfied \cite{Cline:2018ami}.  (The uncertainty in the bound
is due to the unknown 
nuclear equation of state.)  This scenario is highly constrained by a
number of observables sensitive to $\gamma'$, including big bang
nucleosynthesis (BBN), the cosmic microwave background (CMB), 
DM direct detection, supernovae,
and structure formation effects from DM self-interactions.

\begin{figure*}
\centerline{\includegraphics[width=0.45\textwidth]{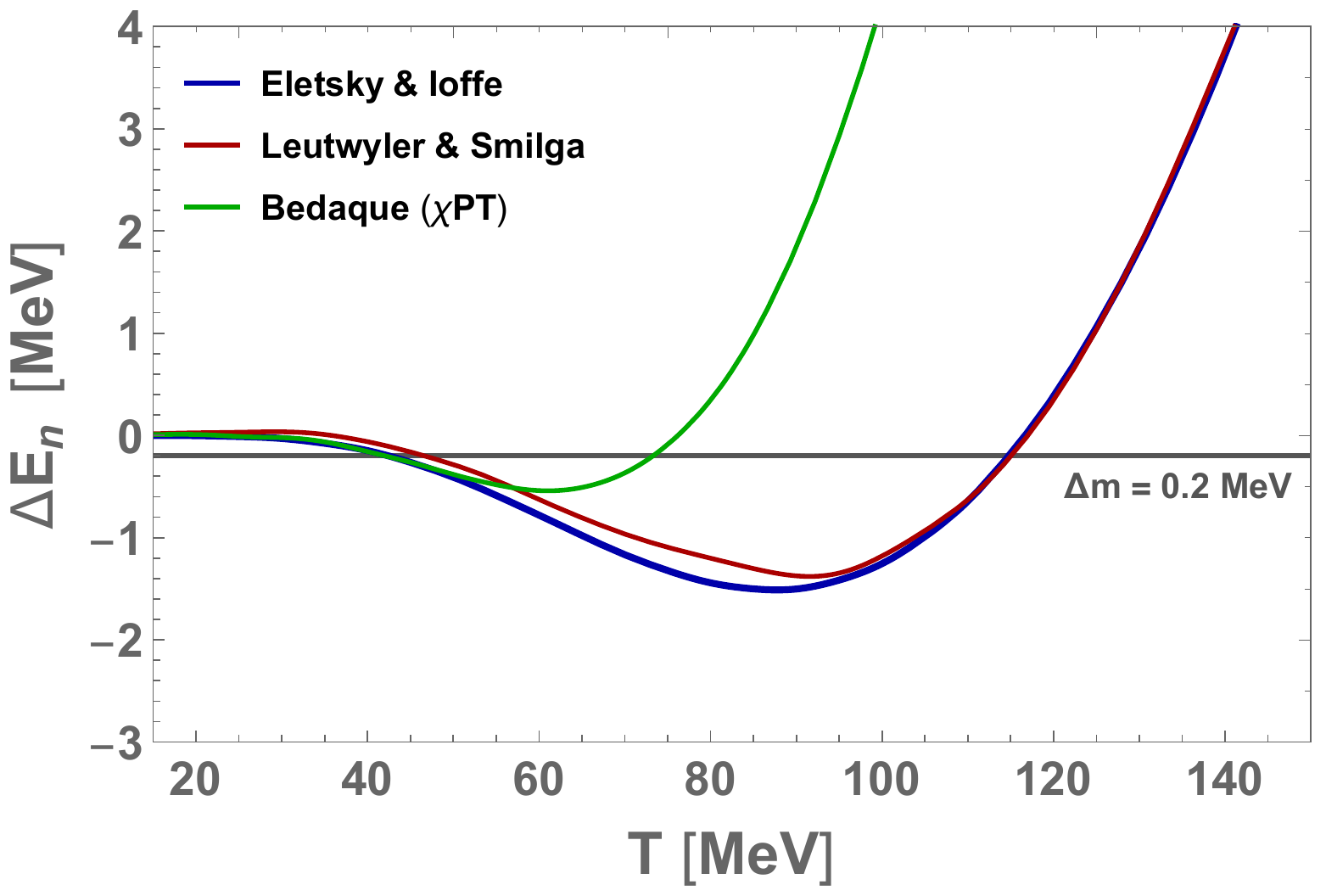}
\hfil\includegraphics[width=0.45\textwidth]{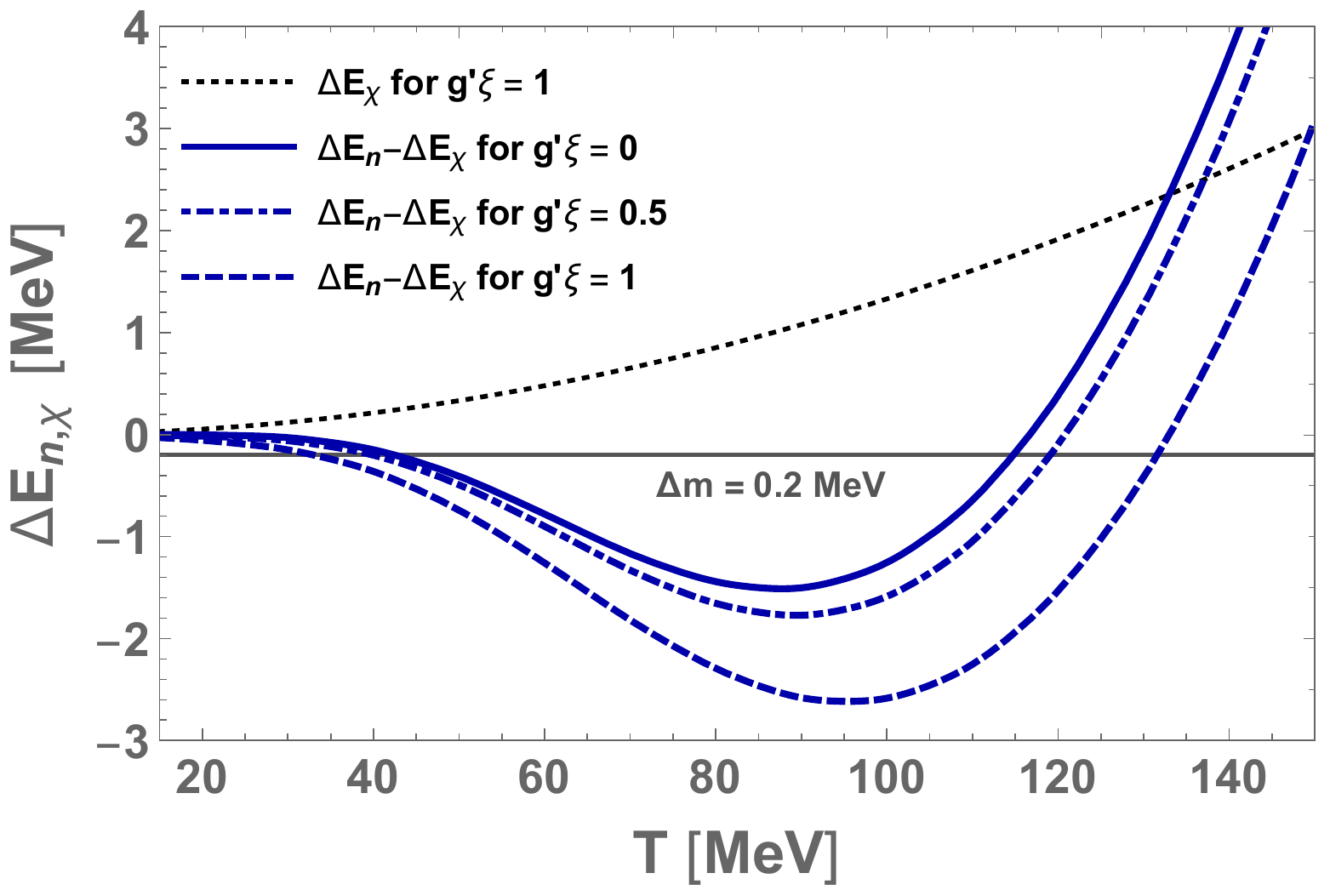}}
\caption{{\it Left panel.} Thermal shifts $\Delta E_n$ to neutron mass
from dispersion relations \cite{Leutwyler:1990uq,
Eletsky:1997dw} (purple, blue) and chiral
perturbation theory \cite{Bedaque:1995pa} (green).  
{\it Right panel.} Effect of including the DM thermal mass.
The dotted curve is $\Delta E_\chi$ for
$g'\xi = 1$ from Eq.~(\ref{Echi}).
Lower curves show the {\it difference} in the thermal self-energies,  
$\Delta E_n - \Delta E_\chi$, for $g'\xi = 0,\,0.5,\,1$.   
For comparison we also indicate the maximal mass difference
$\Delta m\equiv m_n-m_\chi\lesssim0.2$\,MeV compatible with baryon asymmetry
generation and stability of $^9$Be and $^{11}$Be through $n\to\chi+\gamma$;
resonantly enhanced oscillations occur for $\Delta E_n - \Delta E_\chi\cong -\Delta m_\chi$.}
\label{fig:NSE}
\end{figure*}  

In the present work we will show that it is possible to explain both
the baryon asymmetry and the neutron decay anomaly within the same
model, while at the same time evading all cosmological bounds,
if $m_{\gamma'} \sim 60\,$keV and an additional species of dark
radiation $\nu'$ is introduced to enable the decay
$\gamma'\to\nu'\bar\nu'$.  This entails moderate tension with 
BBN and CMB limits on extra radiation species, implying that it 
may be testable already in the near future.  
Alternatively, larger $m_{\gamma'}$ consistent with $\gamma'\to e^+e^-$,
and hence without the need of introducing the additional degrees of freedom
contributed by $\nu'$, 
can be accommodated if we disregard the neutron lifetime anomaly.

\section{Theoretical framework}
\label{sec:framework}

We consider the same model as in Ref.~\cite{Cline:2018ami},
with an elementary Dirac DM particle $\chi$ that carries baryon
number and mixes
with the neutron through the mass term (\ref{Lmix}), and 
a dark photon coupling to $\chi$ with strength $g'$.  
In order to kinematically allow for the decay $n\to \chi\gamma'$,
the $n$-$\chi$ mass splitting must satisfy
\be
	\Delta m\equiv m_n-m_\chi > m_{\gamma'}\,. 
	\label{decaykin}
\ee
However, there is also a subdominant decay channel $n\to \chi\gamma$
in this case, suppressed by the dipole moment $\mu_n$ of the neutron.
This imposes 
\be
	\Delta m <1.665\,{\rm MeV}  
\ee
for stability of $^9$Be, which is the most constraining nuclear
decay \cite{Fornal:2018eol}. We note that $^{11}$Be has an
even smaller neutron binding energy, 0.504\,MeV rather than
1.665\,MeV, which would further strengthen this 
limit~\cite{Ejiri:2018dun}.

Another relevant
parameter is the ratio of the temperatures in the two sectors,
\be
	\xi = {T_{\gamma'}\over T}
\ee
(or $T_{\nu'}/T$ at later times where $\gamma'\to\nu'\bar\nu'$ has
already occurred).  We will show how $\xi<1$ is determined by the
decoupling temperature between the two sectors, which turns out to be
independent of details of the UV physics. 
The baryon asymmetry depends mainly on $\Delta m$
and  the mixing mass
$\delta m$, with weak dependence on the combination $g'\xi$.  

The dark photon mass must satisfy Eq.~(\ref{decaykin}) to explain
the neutron lifetime anomaly.  We will further show that $\Delta
m \lesssim 0.2\,$ MeV is required to simultaneously respect the limit from $n\to\chi\gamma$
and to obtain the observed baryon asymmetry. 
This bound kinematically
forbids $\gamma'\to e^+e^-$ decays, which in turn makes the 
dark photon
longlived, such that it overcloses the 
universe~\cite{Redondo:2008ec,Cline:2018ami}.
To avoid this, we supplement the
model with an inert light particle, dark radiation, taken to be a 
massless Dirac neutrino $\nu'$ that couples to $\gamma'$ with 
strength $Q'_{\nu'} g'$ and charge $Q'_{\nu'}\neq1$.  
This is anomaly-free and allows for  $\gamma'\to \nu'\bar\nu'$, 
while the charge
difference between $\nu'$ and $\chi$ forbids mass mixing 
between the states, that would
lead to proton decay.

A further requirement is that the symmetric component of the DM
must be sufficiently diluted by annihilations.  The
channels $\bar\chi\chi\to \gamma'\gamma',\,\nu'\bar\nu'$ turn out to
be too inefficient since we need a small gauge coupling $g'\sim
10^{-3}$ to satisfy previous requirements.  This can be overcome
by introducing a heavy dark fermion $\psi$ with coupling
$\bar\chi\phi\psi$ to the DM and the dark Higgs $\phi$, allowing for
$\chi\bar\chi\to \phi\phi^*$ through exchange of $\psi$. 

Finally, one should avoid leakage of the primordial $\chi$ asymmetry
into the SM sector at high temperatures, since this would generally
produce too large a baryon asymmetry relative to the DM abundance.
We point out a simple mechanism for enforcing this requirement in the
present model, by adding a very weakly coupled interaction 
$\bar\chi\phi\psi'$ involving a heavy Majorana dark sector fermion
$\psi'$.

\section{$\chi$-$n$ oscillations}
\label{sec:osc}

The oscillations of $\chi$ into neutrons are determined by a $2\times
2$ Hamiltonian that includes the mass terms  and thermal self-energy
corrections from the forward scattering of the fermions on particles
in the plasma.  The most important such interactions are the elastic
scattering of neutrons on pions and  DM on dark photons.

The Hamiltonian is given by
\be
	{\cal H} = \left(\begin{array}{cc} 
	\Delta E_n + m_n & \delta m\\
		\delta m & \Delta E_\chi + m_\chi\end{array}\right)\,,
\label{eq:ham}		
\ee
where $\delta m$ is the mass mixing introduced in Eq.~(\ref{Lmix}).
The thermal energy correction $\Delta E_n$ to the neutron has been 
calculated in Refs.~\cite{Leutwyler:1990uq,Bedaque:1995pa,Dominguez:1992vg,
Kacir:1995gy,Eletsky:1997dw} using various techniques.  Some of these
results are plotted in Fig.~\ref{fig:NSE}\,(left panel).
  The most
reliable ones use dispersion relations, supplemented by experimentally
measured scattering cross sections 
\cite{Leutwyler:1990uq,Eletsky:1997dw}.  We have digitized the results
of Ref.~\cite{Eletsky:1997dw} as our estimate of $\Delta E_n$.  
Using thermal field theory techniques similar to Ref.~\cite{Notzold:1987ik} 
(see appendix {\ref{appA}} for details), we find for
the DM energy shift
\be
	\Delta E_\chi = {g'}^2 {T_{\gamma'}^2\over 8 m_\chi}\,.
\label{Echi}
\ee
This result is also plotted, for $g'\xi=1$, in
Fig.~\ref{fig:NSE}\,(right panel).

The combination $\Delta E_n-\Delta E_\chi$ is relevant 
for getting resonantly enhanced oscillations, which occur when 
$\delta E\equiv \Delta m + \Delta E_n  - \Delta E_\chi$ vanishes
and the eigenvalues of ${\cal H}$ become degenerate.
For illustration, we  plot in Fig.~\ref{fig:NSE}\,(right panel)
the difference in thermal self-energies, with $g'\xi = 0,\,0.5,\,1$,
along with a fiducial value of $\Delta m = 0.2\,$MeV.
Clearly there are two resonant temperatures, where 
the curves cross the horizontal line.  We will show that baryon 
production is dominated by the resonance at the lower temperature.

\subsection{Heuristic approach}

To find the efficiency of $\chi\to n$ oscillations for producing the
baryon asymmetry we first use an approximate formalism \cite{Cline:1991zb} 
that has the advantage of being simple and intuitive.
Diagonalizing ${\cal H}$ leads to the mixing angle $\theta$, 
\be
	\tan (2\,\theta) = {2\,\delta m \over 
\Delta m + \Delta E_n  - 
\Delta E_\chi} \equiv {2\,\delta m \over \raisebox{-0.15em}{$\delta 
E$}}\,.
\label{angle}
\ee
The difference between the eigenvalues is given by
\be
	|\delta\omega| = \sqrt{(\delta E)^2+ 4\,\delta m^2}
\ee

Imagine starting with a state that is purely $|\chi\rangle$ at
$t=0$.  Evolving it with the Hamiltonian ${\cal H}$ in Eq.~(\ref{eq:ham}) leads to
\bea
	|\psi(t)\rangle &=& e^{-i(\omega_1+\omega_2)t/2}
\left[\,\left(\cos{\delta\omega\over 2}\, t - i\cos 2\theta 
	\sin{\delta\omega\over 2}\,
t\right)|\chi\rangle \right. \nn\\
	&-i& \left.\left(\sin 2\theta \sin{\delta\omega\over 2}\,
t\right)|n\rangle\,\right]\,.
\eea
Therefore the probability to oscillate into a neutron is given by
\be
	P_n(t) = \sin^2 (2\theta)\, \sin^2(\delta\omega\, t/2)\,.
\ee
Because of the large rate of interactions $\Gamma_n$ of neutrons
on heat bath pions, however, there may not be time for a full oscillation.  
In general one should carry out a time average, over the
short time scale $1/\Gamma_n$,
\be
	\bar P_n = \Gamma_n \int_0^\infty dt\, e^{-\Gamma_n t}
	P_n(t) = {2\, \delta m^2\over \delta\omega^2 + \Gamma_n^2}\,,
\label{Pneq}
\ee
where we used $\sin^2 2\theta = \delta m^2/(\delta m^2 +
\delta E^2/4)$.

The rate of production of neutrons via oscillations is then
$\Gamma_{\rm osc} = \bar P_n \Gamma_n$ per $\chi$ particle (ignoring
the much smaller elastic scattering rate $\Gamma_\chi$ of DM on
dark photons and scalars\footnote{The cross section for $\phi\chi$
scattering from the interaction (\ref{psi_int}) is $10^{-7}$\,mb,
saturating  the bound (\ref{psibound}).  For $m_\phi=60\,$MeV, for
example, $n_\phi/n_\pi$ is only $\sim 7$ at the resonance temperature.
See also Appendix~\ref{appA}.}).
The inverse process of $\chi$ production must proceed
with the same rate, 
per neutron, so the number density of DM 
overall decreases as $\dot n_\chi=-\Gamma_{\rm osc}(n_\chi-n_n)$. 
Since baryon number, and hence the total 
number of $\chi$'s plus neutrons is conserved,
the equation determining the fraction $f=n_n/(n_\chi+n_n)$ of 
DM converting to neutrons is thus 
\be
	\dot f = \Gamma_{\rm osc}(1-2f)\,,
\ee
which has the solution
\bea
	f &=& \sfrac12\left(1 - \exp\left(-2\int dt\, \Gamma_{\rm
osc}\right)\right)\nn\\
	&\cong& \sfrac12\left(1 -\exp\left(-2\int {dT \over T}\, 
	{\Gamma_n \bar P_n \over H}\right)\right)
\label{feq}
\eea
(with $H\cong1.66\sqrt{g_*}\, T^2/M_p$, 
$M_p = 1.22\times 10^{19}\,$GeV and $g_* \cong 10.75$) 
so long as $P_{\rm osc} < 1$.  We need 
\be
f \cong 0.16
\label{eq:Y_ratio}
\ee
to get the desired abundance ratio $\Omega_\chi/\Omega_B = (1-f)/f
 \cong 5.3$, dictated by the coincidence that $m_\chi\cong m_n$.

It remains to determine the rate $\Gamma_n$.   The cross section 
for neutrons to scatter on pions at low energy 
is~\cite{Fettes:1998ud,RuizdeElvira:2017stg}\footnote{We average
the contributions from the $I=1/2$ and $3/2$ isospin scattering 
lengths as
$\sigma = 4\pi(a_{1/2}^2 + 2\, a_{3/2}^2)/3$.
} 
\be
	\sigma_{n\pi} = 4\pi a_0^2 \cong {0.1\over m_\pi^2}
	\cong 2\,{\rm mb}\,,
\ee
giving the rate 
\be
	\Gamma_n = n_\pi \langle \sigma_{n\pi} v\rangle
\ee
with $\langle{v}\rangle=\sqrt{8T/(\pi m_\pi)} \cong 1.6\sqrt{T/m_\pi}$, where $n_\pi$
is the thermal density of pions, including a factor of 3 for
isospin multiplicity.

\begin{figure}
\centerline{\includegraphics[width=0.45\textwidth]{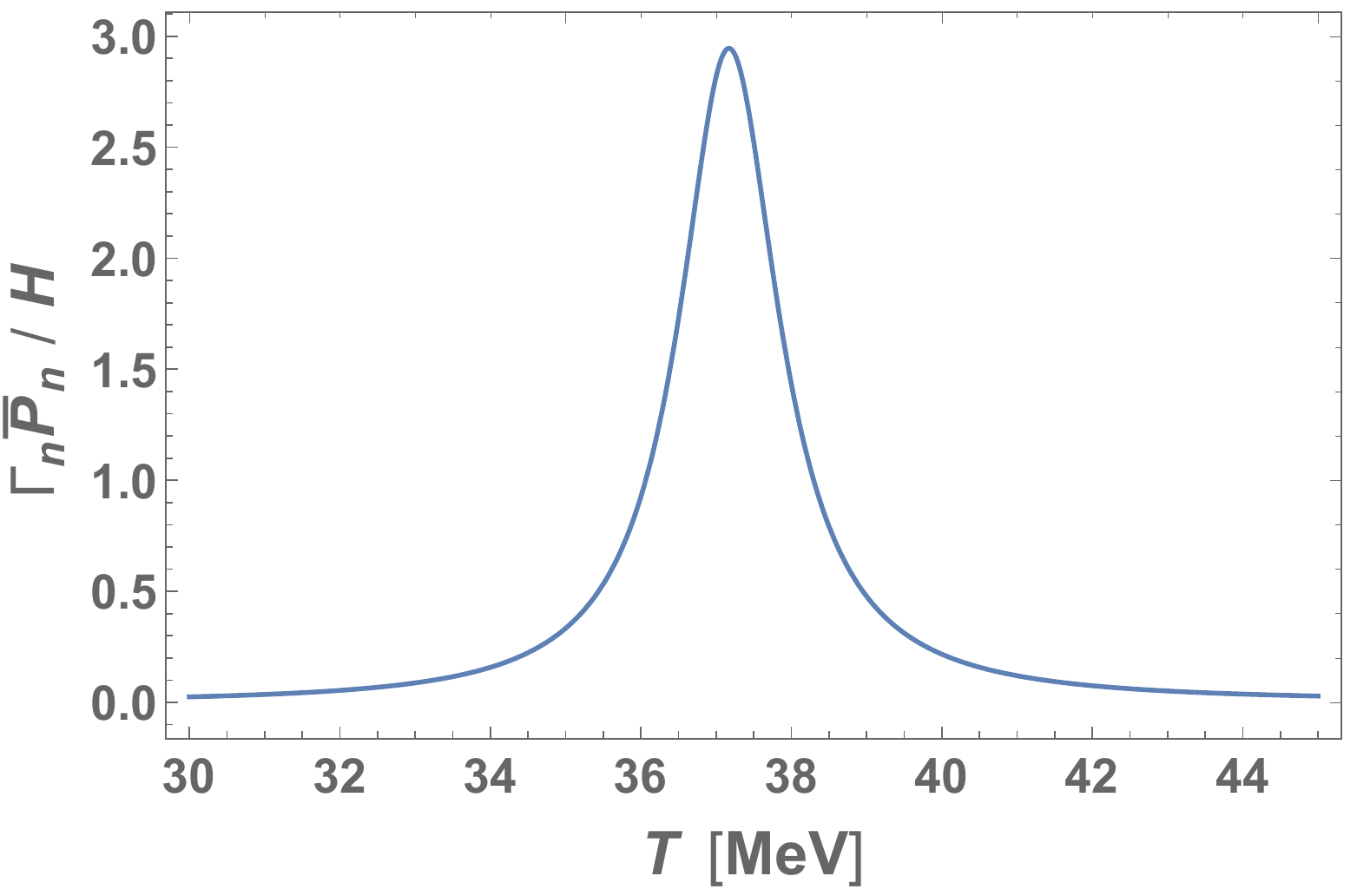}}
\caption{The integrand $\Gamma_n \bar P_n/ H$ of Eq.~(\ref{feq})
versus the photon temperature $T$, showing how baryon number 
generation is dominated by the resonant enhancement of $\chi$-$n$ 
oscillations, using the fiducial parameters $\Delta m = 0.105\,{\rm MeV}$
and $\delta m = 1.05\times10^{-10}\,{\rm MeV}$ of Eq.~(\ref{benchmark}).}
\label{fig:res}
\end{figure}  

This formalism makes it transparent that baryon production is dominated
by resonant enhancement of the oscillations by the finite-$T$
contributions to $\delta E$, which allow for $\delta E(T_r)=0$
at the resonance temperature $T_r$.  At $T=0$, 
tight constraints on $\Delta m$ and $\delta m$ (to be discussed below)
imply a vacuum mixing angle $\theta = 2\delta m/\Delta m$ 
that is by far too small to allow  oscillations 
efficient enough for baryogenesis,
but at $T_r$ the mixing is maximal,
although the oscillation probability is reduced by the damping from
$\Gamma_n$ in Eq.~(\ref{Pneq}).
This is illustrated in Fig.~\ref{fig:res} for a benchmark model, 
Eq.\ (\ref{benchmark}) below,
for which $T_r\cong 37$\,MeV.  Because of the damping, the final
efficiency for baryon production scales as $\delta m^2/\Gamma_n$.

\subsection{Boltzmann equations}
A more rigorous approach is to solve Boltzmann equations for the
relative abundances, including off-diagonal terms arising from
the density matrix, that take into account the oscillations.
The formalism is described for oscillations of DM and
its antiparticle in Ref.~\cite{Cirelli:2011ac,Tulin:2012re}, and requires only
small modifications for neutron-DM oscillations.  In terms
of the independent variable $x \equiv m_n/T$, the equation for the matrix
of abundances $Y=n/s$ is\footnote{Here $s$ is taken to be the
entropy in the visible sector.  Since baryon number $Y_1+Y_2$ is
conserved, we are only concerned with the fraction $f$ of DM that 
converts to baryons, from which $s$ divides out.} 
\bea
	{dY\over dx} &=& -{i\over H x}\left[{\cal H},Y\right]
	-{\Gamma_n\over 2Hx} \left[P_n,\left[P_n,
Y\right]\right] -{\Gamma_\chi\over 2Hx} \left[P_\chi,
	\left[P_\chi,
Y\right]\right]\nn\\
&=&	-{i\over H x}\left(\begin{array}{cc} -\delta m\, Y_-& 
	-\delta m\,\delta Y
	+\delta E\, Y_{12}\\
	\delta m\,\delta Y - \delta E\, Y_{21} & \delta m\, Y_-
	\end{array}\right)\nn\\
	 && -{\Gamma_n+\Gamma_\chi\over 2 H x}\left(
	\begin{array}{cc} 0 & Y_{12}\\ Y_{21} & 0 \end{array}
	\right) \,,
\eea
where we defined $\delta Y \equiv Y_{11}-Y_{22}=Y_n-Y_\chi$, $Y_\pm = Y_{12}\pm
Y_{21}$, and $P_{n,\chi}$ projects onto the state 1 (the neutron)
or 2 (the $\chi$), respectively.  As before, we will ignore
$\Gamma_\chi$ since it is much smaller than $\Gamma_n$.\footnote{It is
worth pointing out that the elastic scatterings on $n$ and $\chi$ 
enter equally into the damping of the oscillations, regardless of
whether one considers $n\to\chi$ or $\chi\to n$ transitions.  This is
because either interaction can measure the state of the system,
regardless of the outcome of the measurement.}

Since $\chi$ can be assigned a baryon number, which is conserved,
the combination $Y_{11} + Y_{22}=Y_n+Y_\chi$ does not evolve.  It is 
then convenient
to recast the equations in terms of $\delta Y$, $Y_+$ and $\eta_-
\equiv i\,\delta m\,Y_-$, 
\bea
	\delta Y' &=& {2\over Hx} \eta_-\,,\nn\\
	\eta_-' &=& {\delta m\over Hx}\left( -2\delta m\,\delta Y
	+\delta E\, Y_+\right) - {\Gamma_n\over 2 H x}\eta_-\,,\nn\\
	Y_+' &=& -{\delta E\over \delta m\, Hx} \eta_- - 
	{\Gamma_n\over 2 H x}Y_+\,.
\label{Beq}
\eea
This form has the advantage that all the functions are now
real-valued, and the coefficients are of order unity, for cases of
interest.  The rescaling of $Y_-$ by $\delta m$ alleviates
the problem of stiffness in the numerical integration.
We integrate the system (\ref{Beq}) with initial 
conditions $Y_{22} = Y_\chi^\infty$ 
and $Y_{ij}=0$ for the other components.  Since
$Y_{11}+Y_{22}$ is conserved, the fraction of DM that converts to 
neutrons is $f = Y_{11}/Y_\chi^\infty = 1-Y_{22}/Y_\chi^\infty$.  
We note that this fraction is necessarily independent of the initial
DM abundance  $Y_\chi^\infty$, so for any value of $f$ we can simply 
rescale $Y_\chi^\infty$ such that the observed DM density 
$\rho_{\rm CDM}=sm_\chi Y_{11}^0$ is 
recovered at late times (in practice, we use $Y_\chi^\infty=1$
as initial condition).
Given that the DM and neutron are
nearly degenerate,  we can identify $f/(1-f) = \Omega_b/\Omega_{\rm
CDM}$ at late times.

\subsection{Results}

\begin{figure}
\centerline{\includegraphics[width=0.5\textwidth]{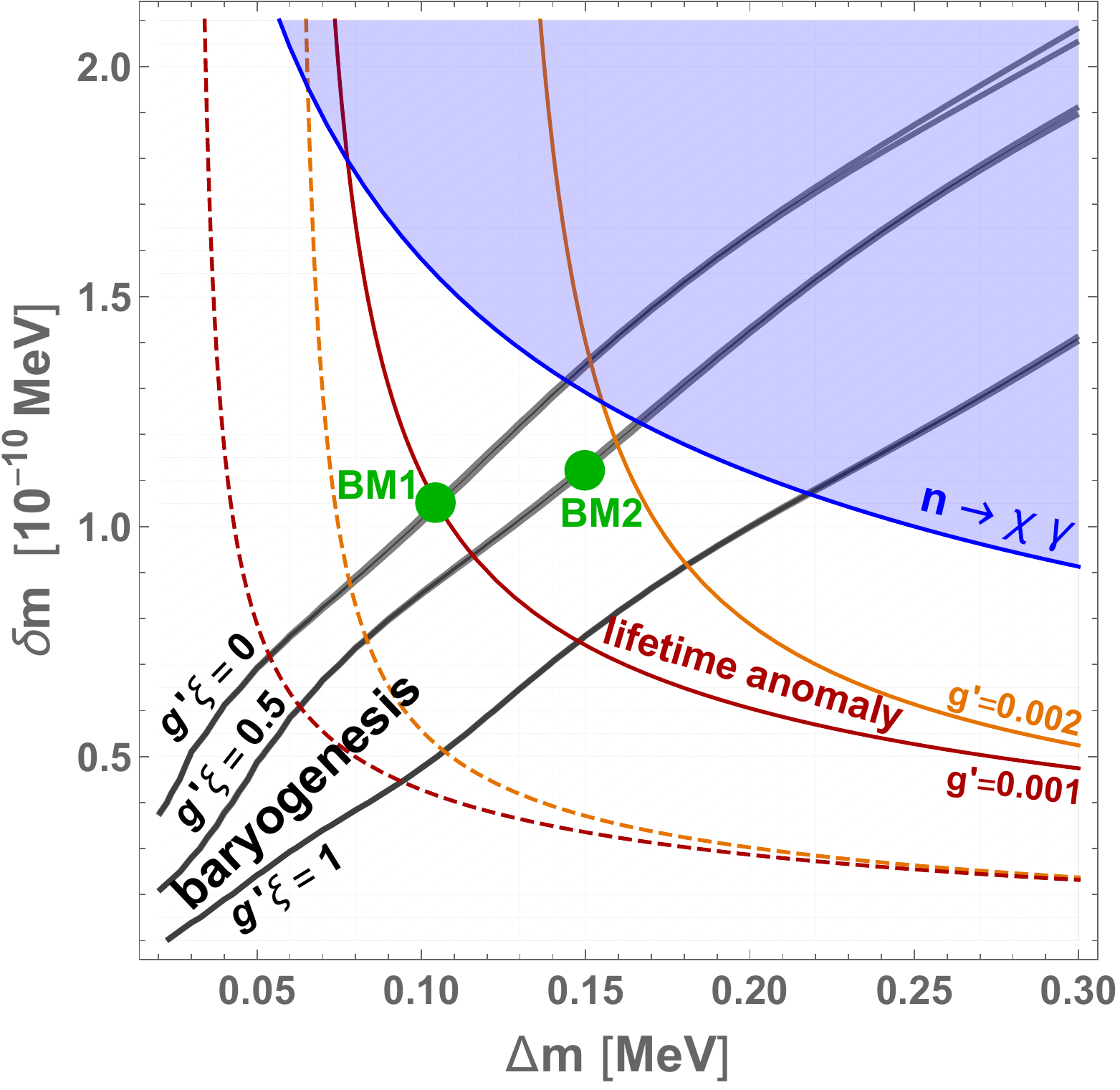}}
\caption{Black contours give the right baryon asymmetry
as in Eq.~\ref{eq:Y_ratio}), in
the $\delta m$-$\Delta m$ plane, for several values of $g'\xi$.
The differences between the baryogenesis predictions from the 
matrix Boltzmann equations
(\ref{Beq}) and the heuristic
method given by Eq.~(\ref{feq}) are not visible on this scale. 
The blue area is excluded from $n\to \chi\gamma$ decays
(stability of $^9$Be).
Red (orange) solid lines indicate the values of $\delta m$ needed to 
explain the neutron lifetime anomaly, assuming 
$m_{\gamma'}/g' = 60\,$MeV and  $g' = 0.002$ ($0.001$).
The lower dashed curves are the same but assuming 
$m_{\gamma'}/g' = 30\,$MeV.}
\label{fig:cont}
\end{figure}  

Fig.\ \ref{fig:cont} shows the correlation between $\delta m$ and
$\Delta m$ needed to get the observed baryon asymmetry, 
by fixing $f$ as stated in Eq.~(\ref{eq:Y_ratio}), for several
values of $g' \xi$.  Each curve is doubled, with a small splitting,
showing the good agreement between the heuristic treatment (lower)
and Boltzmann equation (upper).  We noted above that the baryon
asymmetry scales as $\delta m^2/\Gamma_n$.
For larger $\Delta m$, the resonance temperature is increased, and the
rate of $\Gamma_n$ is larger due to the higher density of pions, 
explaining why $\delta m$ must increase
with $\Delta m$ to keep the baryon asymmetry constant.

On the same figure we plot the upper limit on the vacuum mixing 
angle from null searches for the decay $n\to\chi\gamma$ (blue curve), 
whose branching
ratio must be $\lesssim 0.17$ times the 0.9\% needed to resolve the
lifetime discrepancy \cite{Tang:2018eln}.  This gives
\be
	\delta m \lesssim 5\times 10^{-11}\,{\rm MeV}\left(1\,{\rm
MeV}\over \Delta m\right)^{1/2}
\ee
Fig.~\ref{fig:cont} thus implies that for reasonable values of $g'\xi$,
the baryon asymmetry can be explained only if $\Delta m\lesssim
0.2$\,MeV.

We next consider the neutron lifetime
anomaly, for which  the mixing mass should satisfy \cite{Fornal:2018eol}
\be
	\delta m = {4.2\times 10^{-13}\over 
	\left(1 - (m_{\gamma'}/\Delta
m)^2\right)^{3/4}}\,\left({m_{\gamma'}\over g'}\right) \left(1\,{\rm
MeV}\over \Delta m\right)^{1/2}\,.
\label{lifetime}
\ee
Combining this with the limit on $m_{\gamma'}/ g'$ from neutron stars 
in Eq.~(\ref{NS}), this becomes a constraint on $\delta m$ as a
function  of $\Delta m$ and $g'$ (with a slight dependence upon the
assumed  nuclear equation of state).  Using $m_{\gamma'}/ g'=60$\,MeV,
we  indicate in Fig.~\ref{fig:cont} the values of $\delta m$ that are
required  to resolve the lifetime anomaly for two representative
values of $g' = 0.001,\,0.002$ (solid red and orange curves,
respectively).   It is possible to choose a smaller value of
$m_{\gamma'}/ g'$ (which is consistent with the neutron star
constraint \ref{NS}), 
leading to the lifetime anomaly being resolved by smaller values of $\delta m$.
The choice $m_{\gamma'}/ g' = 30$\,MeV is plotted for illustration
(dashed red and orange curves) for the same two values of $g'$.
This value of $m_{\gamma'}/ g' = 30$\,MeV is still consistent with 
constraints on the DM self-interaction rate; see 
Section \ref{sec:sub} below.

To find a consistent intersection of the baryon asymmetry and
lifetime anomaly curves below the $n\to\chi\gamma$ constraint
in Fig.~\ref{fig:cont},
it is necessary to take $g'\lesssim0.002$. 
The neutron star constraint in Eq.~(\ref{NS}) hence implies that 
the dark photon must be very light, $m_{\gamma'}\lesssim 100$\,keV.
Since $\xi < 1$, the thermal self-energy of $\chi$ as given in 
Eq.~(\ref{Echi}) is furthermore irrelevant; this effectively places 
us on the baryon asymmetry curve corresponding to $g'\xi=0$.
As an example, the benchmark values 
\bea
	\hbox{\bf B1}~~~~\Delta m &=& 0.105\,{\rm MeV}, \quad
	\delta m = 1.05\times10^{-10}\,{\rm MeV}\nn\\
	g' &=&0.001, \quad m_{\gamma'} = 60\,{\rm keV}
\label{benchmark}
\eea
are compatible with all the constraints ($\xi$ being in principle
unconstrained but, as we will see shortly, uniquely determined
by the decoupling temperature).  

The dark photon mass in these scenarios is 
significantly below the $2 m_e$ threshold. In the absence of 
additional channels the dominant decay mode would then be
$\gamma'\to 3\gamma$, which is so slow that the initially thermally 
distributed dark photons would be metastable and overclose the universe \cite{Redondo:2008ec}
(see Ref.~\cite{Cline:2018ami} for additional constraints).
This motivates us to introduce the additional massless 
Dirac neutrino $\nu'$ already mentioned in Section \ref{sec:framework}, 
enabling the much more efficient decay  $\gamma'\to\nu'\bar\nu'$.

Alternatively, if we ignore the neutron decay anomaly (assuming, for
example, that it is due to experimental error)  and insist only on
low-scale baryogenesis, it is not necessary for
$g'$ to be small, nor for the dark photon mass to respect the kinematical
constraint (\ref{decaykin}).  This in turn allows heavier dark photons and
the possiblity for fast $\gamma'\to e^+e^-$ decays through
kinetic mixing, via the Lagrangian term 
$-(\epsilon/2) \, F_{\mu \nu} F'^{\mu \nu}$,
without the need to  introduce
additional light degrees of freedom like $\nu'$.\footnote{Some extra
field carrying U(1)$'$ charge will be needed to insure U(1)$'$ charge
neutrality of the universe at early times, as we will discuss in 
Section \ref{sec:seq}.   However it need not be light, and so will
not necessarily contribute extra radiation degrees of freedom that
contribute to $N_{\rm eff}$. See appendix \ref{sec:numass} for more details.}

In this simpler scenario, the 
additional heavy fermion  $\psi$ described in Sec.\ 
\ref{sec:framework} is also no longer required,
since with sizeable values of $g'$ the process
$\chi\bar\chi\to\gamma'\gamma'$ can efficiently 
annihilate away the symmetric DM component. Possible benchmark 
values for such a minimal baryogenesis scenario are
\bea
	\hbox{\bf B2}~~~~\Delta m &=& 0.15\,{\rm MeV}, \quad
	\delta m = 1.12\times10^{-10}\,{\rm MeV}\nn\\
	g'\xi &=& 0.5, \quad m_{\gamma'} = 50\,{\rm MeV}.
\label{benchmark2}
\eea

In the {\bf B2} model, an upper bound on the kinetic mixing
parameter $\epsilon$ comes from the
scattering interactions of $\chi$ off protons,  with a cross section
\be
	\sigma_{\chi p} = {4\alpha(g'\epsilon)^2\, \mu_{p\chi}^2\over 
m_{\gamma'}^4}
\ee
where $\mu_{p\chi}$ is the reduced mass of $\chi$ and the proton. 
Results from CRESST-III limit 
$g' \epsilon \lesssim 1.2 \times 10^{-7}$~\cite{Petricca:2017zdp}.  
Even more constraining are the beam dump experiments at 
Orsay~\cite{Davier:1989wz} and SLAC (E137)~\cite{Bjorken:1988as}, 
which limit $\epsilon \gtrsim 1.5 \times 10^{-5}$ or 
$\epsilon \lesssim 4.2 \times 10^{-8}$ \cite{Andreas:2012mt}, and 
observations of the flux of neutrinos from supernova 1987A, which 
constrain $\epsilon \gtrsim 4.7 \times 10^{-8}$ or 
$\epsilon \leq 3.2 \times 10^{-10}$~\cite{Chang:2016ntp}.

The decay of the dark photon to electrons and positrons proceeds with a width and lifetime given by
\bea
	\Gamma &=& {{\alpha\epsilon^2}\over 3\, m_{\gamma'}}
	\left(m_{\gamma'}^2 + 2 m_e^2\right)\left(1 - 4
m_e^2/m_{\gamma'}^2\right)^{1/2}\nn\\
	\tau &\cong & 0.5\left(10^{-10}\over\epsilon\right)^2\,{\rm s} \, .
\eea
The above constraints combine to limit $\tau \gtrsim 0.05$\,s, 
which is sufficiently early 
to avoid disturbing the production of light elements during BBN \cite{Hufnagel:2018bjp,Forestell:2018txr}. 
As will be discussed 
in section \ref{sec:dof}, interactions other than the kinetic mixing can 
thermalize the dark and visible sectors, so there is no constraint on 
$\epsilon$ from this requirement.

\section{Thermal history}

To have a fully consistent scenario, we need to address several
issues: (1) redistribution of the presumed initial $\chi$ asymmetry
into baryons should not have taken place at a higher temperature 
through the interactions present in our model, since this might be
more important than the $\chi$-$n$ oscillations; (2) the
U(1)$'$-breaking transition where $\phi$ gets its VEV must occur
before the oscillations go through resonance; 
(3) the additional
light degrees of freedom in the dark sector must be compatible with
constraints from BBN and CMB; (4) the symmetric component of the
$\chi$ relic density must be sufficiently suppressed, to avoid having
more DM  than is observed; (5) the self-interaction cross
section of $\chi$ should respect constraints from structure
formation.  We consider these in turn.
The constraints (2--5) are automatically  satisfied for the 
benchmark point {\bf B2} if they are fulfilled for {\bf B1}, so we 
adopt the {\bf B1} parameters in this section.

\subsection{Sequestering of $B$ at high $T$}
\label{sec:seq}

Our assumption is that initial asymmetries in $\chi$, $\phi$, and
$\nu'$ were somehow created at a high temperature, while the standard
model baryon and lepton asymmetries were initially vanishing. 
Asymmetries in $\phi$ or $\nu'$ are needed in addition to that in
$\chi$ to maintain U(1)$'$ charge neutrality of the universe, since
U(1)$'$  remains unbroken until $T\sim 100\,$MeV (see section
\ref{breaking}). However the UV model of reference
\cite{Cline:2018ami}, augmented for the present purposes (see section
\ref{decoupling}), includes interactions which at high
temperatures will redistribute these initial asymmetries into the
other dark sector particles as well as into the visible sector. The
relevant interactions are 
\bea
\label{eq:LagUV}
	\mathcal{L}_{\rm UV} &=& \lambda_1 \bar d^a P_L \chi \Phi_{1,a} 
	+ \lambda_2 \epsilon^{abc}\bar u^{\sss C}_a P_R d_b \Phi_{2,c}	\nn\\
	&+& \lambda_3\bar\chi\phi\psi + 
	\mu\, \phi\, \Phi_{1,a}\Phi^{*a}_2\ + {\rm H.c.}\, .
\eea
$\lambda_{1,2,3}$ are dimensionless coupling constants, while $\mu$ is a
constant with units of mass. 
$\Phi_1$ and $\Phi_2$ are both TeV scale
scalar fields that are colored triplets with baryon number $-2/3$, and
$\Phi_1$ also caries U(1)$'$ charge. 
The resulting decays and inverse decays will be in equilibrium at
high temperatures, redistributing any initial asymmetry amongst all
particles in the plasma.
Generically, this will induce an early baryon asymmetry of the same order
as that in DM, contrary to 
our hypothesis that only neutron-DM oscillations are important.

To see this, it is sufficient to consider temperatures just below the
electroweak phase transition (EWPT), $T_c\cong 160\,$GeV, at which the relevant SM interactions 
are sphalerons (still in equilibrium since the EWPT is a cross-over
transition) and $W^-\leftrightarrow d\bar u$, $W^-\leftrightarrow
e\bar\nu$. Following Ref.~\cite{Harvey:1990qw}
we take the 
up- and down-type quarks of all generations to have 
common chemical potentials $\mu_u$ and $\mu_d$, and denote the
sum of neutrino chemical potentials by $\mu_\nu$.  (The charged lepton
potentials $\mu_L$ are eliminated in favor of $\mu_W$ and $\mu$.)  
At this scale the effective operator from integrating out $\Phi_i$
is
\be
	\frac{\lambda_1 \lambda_2 \mu}{m^2_{\Phi_1} m^2_{\Phi_2}}
 \phi \,\epsilon^{a b c} \left( \bar u^{\sss C}_a P_R d_b \right) 
\left( \bar \chi P_R d_c \right) \, 
\label{lowEint}
\ee
which is in equilibrium down to $T \cong 11$\,GeV for a light dark scalar
$\phi$ (see appendix \ref{dim7freezeout}).\footnote{After this interaction no longer keeps $\chi$ in equilibrium, it will lead to $\chi$ decays before the QCD phase transition. However, the lifetime for such a decay is $\sim 10$ seconds for our benchmark points, so the effect of these decays on the abundance of $\chi$ particles will be minimal.}  The coefficient of (\ref{lowEint}) is fixed,
since it gives rise to the $n$-$\chi$ mass mixing term
when $\phi$ gets its VEV, 
\be
	\label{eq:deltam}
	\delta m = {\lambda_1\lambda_2\beta \langle\phi\rangle \mu\over
	m_{\Phi_1}^2 m_{\Phi_2}^2} \sim 10^{-10}\,{\rm MeV}\, .
\ee
where $\beta = 0.014\,$GeV$^3$ \cite{Aoki:2017puj} from the lattice
matrix element $\langle n|udd|0\rangle$. 

The equilibrium constraints from Eq.~(\ref{lowEint}) and the SM 
interactions~\cite{Harvey:1990qw} are then
\bea
	\mu_\chi &=& \mu_\phi + \mu_u + 2\mu_d\\
	\mu_W &=& \mu_d - \mu_u \label{eq:wdu}\\
	0 &=& 9\mu_u + 6\mu_W + \mu_\nu
\eea
These are supplemented by the vanishing of the conserved charges, electric and
U(1)$'$,
\bea
	Q &\propto& 3\mu_u -\mu_\nu - 9\mu_W = 0\\
	Q' &\propto& \mu_\chi + \mu_\phi + Q'_{\nu'}\mu_{\nu'} = 0
\label{muQnu}
\eea
where we have used that
$Q'_{\nu'}$ is the $\nu'$ charge relative to that of 
$\chi$ or $\phi$.  These equations give quark asymmetries
\be
	\mu_u = \frac1{11}(\mu_\chi-\mu_\phi),\quad
	\mu_d = 5\mu_u
\ee
unless $\mu_\chi = \mu_\phi$ at this temperature. Such a cancellation
does not come about without tuning the relative initial asymmetries of
$\chi$ and $\nu'$. How these chemical potentials are related to the
initial asymmetries, possibly generated at a much higher scale, is discussed in appendix \ref{app:ChemPot}.  The
$\nu'$ asymmetry is determined by Eq.\ (\ref{muQnu}) alone.

However there is a simple low-energy mechanism for enforcing
$\mu_\phi = \mu_\chi$, supposing there is a massive Majorana fermion $\psi'$
in the dark sector with coupling
\be
	\lambda' \bar\chi\phi\psi' + {\rm H.c.}
\label{majorana_int}
\ee
If in equilbrium, this gives the desired relation $\mu_\phi =
\mu_\chi$, since $\mu_{\psi'} = 0$ by the Majorana nature of $\psi'$.
The decays $\psi'\to\chi\phi$ are in equilbrium for $T\ge T_c=160\,$GeV so
long as 
\be	
	\lambda’ \gtrsim 1 \times 10^{-7} 
	\left( \frac{100 \ {\rm GeV}}{m_{\psi’}} \right)^{1/2} .
\label{majorana}
\ee
This supplies the extra condition needed to ensure that no baryon
asymmetry is induced prior to the onset of $n$--$\chi$ oscillations.

The new interaction (\ref{majorana_int}) is potentially dangerous since
it violates baryon number.  The dominant process is the $\Delta B=2$
oscillations of $\chi$--$\bar\chi$ due to the mass term
\be
	\sfrac12{\delta m_\chi}
	\bar\chi\chi^{\sss C} + {\rm H.c.} = {\lambda'^2 v'^2\over 2 m_{\psi'}}
	\bar\chi\chi^{\sss C} + {\rm H.c.}
\ee
arising after $\phi$ gets its VEV $v'/\sqrt{2}$ and integrating out
$\psi'$.  In our scenario,
this can be very small, $\delta m_\chi \cong 3\times 10^{-18}\,$MeV
for $v' = 60\,$MeV as we have assumed, and $\lambda'$ saturating
the bound (\ref{majorana}) with $m_{\psi'}=100\,$GeV.
Such oscillations
have been studied in detail in Ref.~\cite{Tulin:2012re}.  They are
damped by the scattering of $\chi$ on dark photons as long as the
scattering rate $\Gamma_s\sim g'^4 T^3/m_{\gamma'}^2$ exceeds the 
oscillation rate $2 \delta m_\chi$, true for temperatures
\be
   T\gtrsim \left(2\delta m_\chi m_{\gamma'}^2\over g'^4\right)^{1/3}
\sim 3\,{\rm keV}
\label{B2oscil}
\ee
which is far below the scales of interest for baryogenesis through
$\chi$-$n$ oscillations.   The estimate (\ref{B2oscil}) applies for 
the {\bf B1} benchmark model, but is even lower for {\bf B2.}

\subsection{U(1)$'$-breaking phase transition}
\label{breaking}

The dark scalar $\phi$ must get its VEV
before the $n$-$\chi$ oscillations begin, since the mass mixing
term $\delta m$ coming from (\ref{lowEint}) is proportional to
$\langle\phi\rangle$ in our model.
Because the gauge coupling $g'$ is very small, the phase transition
in the dark sector is controlled by the dark scalar $\phi$ alone.
Supposing its potential is
\be
	V_0 = \lambda_\phi(|\phi|^2 - v'^2/2)^2
\label{V0eq}
\ee
the field-dependent masses for the real and imaginary parts of 
$\phi = (\varphi_r+i\varphi_i)/\sqrt{2}$ are 
\bea
	m_r^2&=& {\partial^2 V\over \partial\varphi_r^2} = 
	\lambda_\phi(6|\phi|^2 - v'^2),\nn\\
	m_i^2 &=& {\partial^2 V\over \partial\varphi_i^2} = \lambda_\phi(2|\phi|^2 - v'^2)
\eea
(assuming that $\varphi_r$ is the component that gets the VEV).
At mean-field level, the finite-temperature contribution to the 
potential is \cite{Dolan:1973qd}
\be
	V_T = {T'^2\over 24}(m_r^2 + m_i^2)
\label{VTeq}
\ee
where $T'$ is the temperature of the dark sector.  Combining
(\ref{V0eq}) and (\ref{VTeq}) one finds that the critical temperature,
where the curvature of the full potential at $\phi=0$  vanishes, 
is given by
\be
	T'_c = \sqrt{3}\,v' = \sqrt{3}\,{m_{\gamma'}\over g'}
\cong 104\,{\rm MeV}
\ee
independently of $\lambda_\phi$, so long as $\lambda_\phi \gg g'^2$.
Since $T' = \xi T$ with $\xi<1$, we see that the critical temperature
as measured in the visible sector is even larger, $T_c > 104\,$MeV.
This is comfortably above the resonance temperature $T_3\sim 30$\,MeV
for our benchmark model.

\subsection{Constraints on extra degrees of freedom}
\label{sec:dof}
The dark sector in our model has new light degrees of freedom
at the scale of BBN, the $\sim 60$\,keV dark photon and 
two massless Weyl neutrinos $\nu'$, the latter being also present
during temperatures relevant for the CMB.   We show in appendix
\ref{dim7freezeout} that the effective operator (\ref{lowEint}) keeps
the two sectors in equilbrium until $T \cong 11\,$GeV, irrespective of
the details of the UV model parameters.
This relatively high decoupling temperature helps to reduce the number of
extra radiation degrees of freedom, conventionally parametrized as
the number of additional effective neutrino species, 
$\Delta N_{\rm eff}$,

 Following decoupling, the entropies in
the two sectors are separately conserved.  During the QCD transition,
the relative temperature $\xi=T'/T$ goes down because the number of
SM degrees of freedom, $g_\mathrm{SM}(11\,\mathrm{GeV})\simeq 86.25$, 
decreases by a greater factor than the
corresponding decrease in the dark sector
($\chi+\nu'+\phi+\gamma' \to \nu' + \gamma'$).
We thus expect
\be
	\xi_{BBN} \cong \left(86.25\over 10.75\right)^{-1/3}
\left(6.5\over 11\right)^{-1/3} \cong 0.59
\ee
at $T > m_e$ when electrons are still in equilibrium.  At this
temperature, the dark photons and neutrinos are both present in the
dark sector, and they contribute 
\be
	\Delta N_{\rm eff} = \left({4\over 7}\times 3 +
2\right)\xi_{BBN}^4
	\sim 0.47
\ee
to the effective number of neutrino species.  This is below the 
3$\sigma$ bound $\Delta N_{\rm eff} < 0.54$, but larger than the
2$\sigma$ bound $\Delta N_{\rm eff} < 0.31$
allowed by BBN \cite{Hufnagel:2017dgo}
(similar BBN limits are reported by Cyburt {\it et al.}~\cite{Cyburt:2015mya}).

At lower temperatures, the number of relativistic degrees of freedom
again changes in both sectors, as electrons and dark photons
disappear.  At the time of the CMB, we have a temperature ratio of
\be
	\xi_{CMB} \cong \left({10.75\over 3.91}\times {3.5\over
6.5}\right)^{-1/3} \xi_{BBN} \cong 0.52\,.
\ee
Taking into account that the temperature ratio of SM neutrinos 
and photons has also increased by that time, we hence get
\be
	\Delta N_{\rm eff} = 2\, \xi_{CMB}^4 \times (11/4)^{4/3}
	\sim 0.56\,.
\ee
This is above the 3$\sigma$ limit of $\Delta N_{\rm eff} < 0.45$
from Planck \cite{Aghanim:2018eyx,pla}, combining CMB and BAO
measurements. But we note that 
this limit weakens to $\Delta N_{\rm eff} < 0.55\,(0.66)$ at 
2$\sigma$\,(3$\sigma$)~\cite{pla}
when adding the direct measurement of the Hubble rate, 
$H_0=(73.45 \pm 1.66)$\,km\,s$^{-1}$\,Mpc$^{-1}$ \cite{riess+}, to these 
datasets.

\begin{figure}
\centerline{\includegraphics[width=\columnwidth]{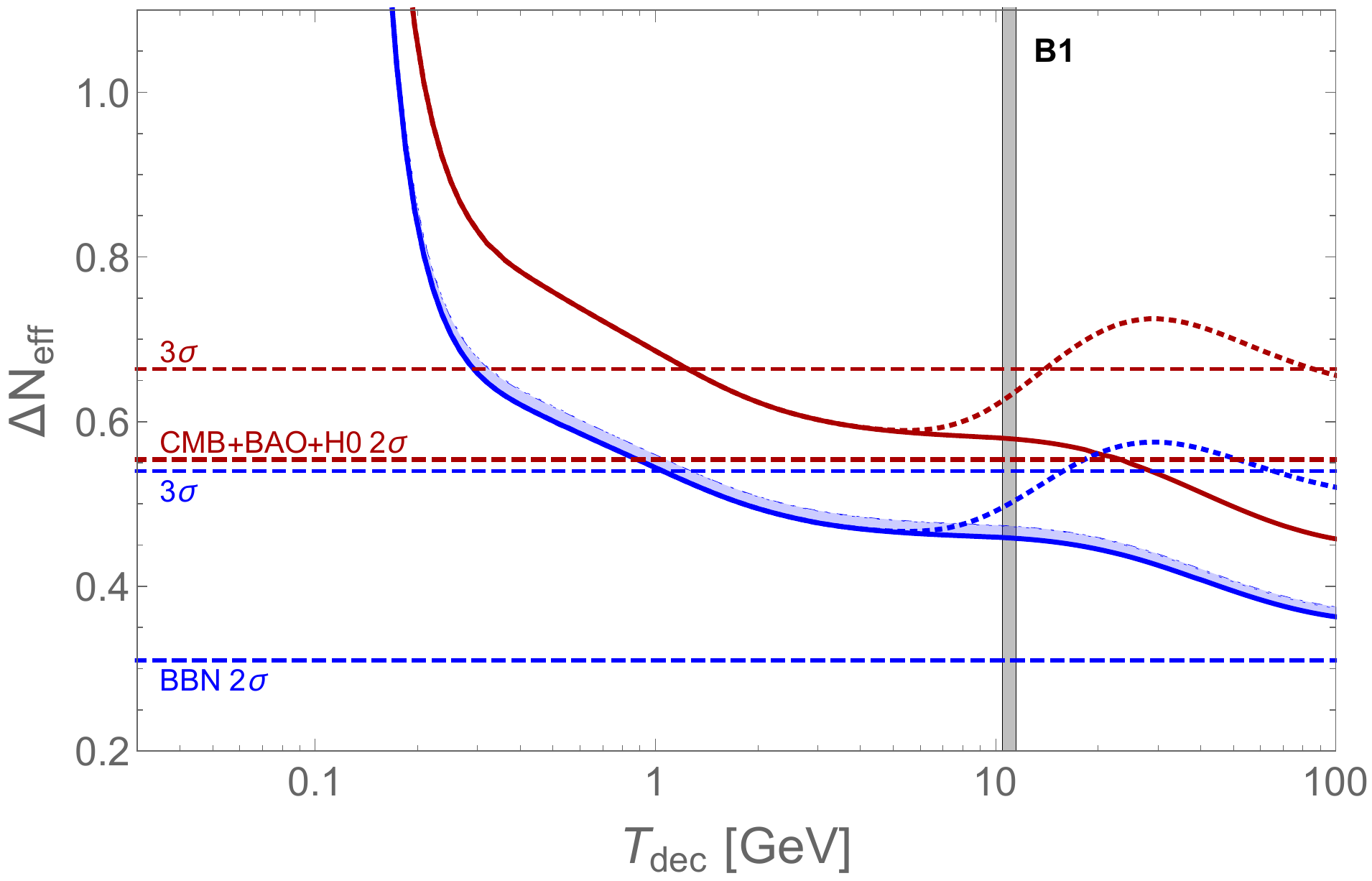}}
\caption{The extra contributions to the effective number of
neutrino species, $\Delta N_{\rm eff}$, from the dark neutrinos
$\nu'$, the dark photon $\gamma'$ (at $T\gtrsim 60\,$keV), and the
dark Higgs boson $\phi$ (for $T\gtrsim 60$\,MeV). 
Dotted lines show the effect of adding a fermion
with $m_\psi=50$\,GeV.
The vertical line indicates the decoupling temperature $T_{\rm dec}$
for the benchmark point {\bf B1}. 
We also indicate limits from BBN \cite{Hufnagel:2017dgo} and 
Planck data \cite{Aghanim:2018eyx}. 
}
\label{fig:Neff}
\end{figure}

For a more quantitative treatment we implemented the full temperature 
dependence of the energy density for each dark sector 
d.o.f.~in \ds~\cite{Bringmann:2018lay}, rather than the relativistic 
limit used in the above estimates, and plot the expected $\Delta N_{\rm eff}$
as a function of decoupling temperature $T_{\rm dec}$ in Fig.~\ref{fig:Neff}.
The red line shows $\Delta N_\mathrm{eff}$ at
CMB times, while the blue line shows this quantity for $T=1$\,MeV; the blue
band corresponds to $\Delta N_\mathrm{eff}$ for the whole range of temperatures
$0.1\,\mathrm{MeV}<T<5$\,MeV most strongly probed by BBN.
We also indicate in this plot the already mentioned constraints from BBN 
and CMB.
For the benchmark point {\bf B1},  Fig.~\ref{fig:Neff} confirms the above 
estimates for $\Delta N_\mathrm{eff}$.
At the same time, this figure illustrates the effect of changing  
$T_{\rm dec}$. For example, for a smaller $\langle\phi\rangle=m_{\gamma'}/g'$
the coefficient in front of Eq.~(\ref{lowEint}) would increase, leading to a 
decrease in $T_{\rm dec}$ that sharpens the tension with BBN and CMB.
A larger value
of $\langle\phi\rangle$, on the other hand, cannot be achieved in view of the 
neutron star constraint (\ref{NS}). To {\it decrease} the tension with $\Delta N_\mathrm{eff}$
would thus require to modify the low-energy operator given in Eq.~(\ref{lowEint})
-- or to make the dark photon and Higgs non-relativistic already during BBN, 
as in benchmark point {\bf B2}. 
Finally, if decoupling happens while some of the heavy degrees of
freedom are still in equilibrium, this will increase the entropy ultimately
dumped into $\nu'$; we illustrate this in Fig.~\ref{fig:Neff} by showing the effect
of an additional fermion with mass $m_\psi=50$\,GeV (dotted lines).

\subsection{Dark matter annihilation and kinetic decoupling}
\label{decoupling}

Although we are assuming that the asymmetric component of the  DM
abundance is generated by an unknown mechanism, the symmetric (thermal)
component is determined by the annihilations  of $\chi\bar\chi$ into
lighter particles.  Because of the small value of $g'=10^{-3}$, the
cross sections for $\chi\bar\chi \to \gamma'\gamma'$ and 
$\chi\bar\chi \to \nu'\bar\nu'$ are inadequate for avoiding overclosure.
However the channel $\chi\bar\chi \to \phi\phi^*$ becomes available
if we introduce a neutral Dirac fermion $\psi$ with coupling
\be
	\lambda_3 \bar\chi\phi\psi + {\rm H.c.}
\label{psi_int}
\ee
This is consistent with stability of $\chi$ as long as $m_\psi >
m_\chi$.  The cross section for $\chi\bar\chi \to \phi\phi^*$ by
$\psi$ exchange, to leading order in the center-of-mass velocity 
$v_{\rm cm}$ of the DM and for $m_{\psi}\gg m_\chi$, is
\be
	\sigma v_{\rm rel} = {\lambda_3^4\, v_{\rm cm}^2\over 16\pi
	m_\psi^2}\,.
\ee
After thermally averaging, we find that to avoid 
overclosure one needs ${m_\psi\lambda_3^{-2}} \lesssim 1\,{\rm TeV}$ \cite{Graesser:2011wi}.

For a more quantitative estimate one needs to take into account the non-standard 
temperature evolution in the dark sector. To this end, we implemented
the exact $\xi(T)$ dependence as well as the full expressions for 
$\sigma v_{\rm rel}$ in \ds. Solving the Boltzmann equation numerically
for input parameters in the range of interest,
we find that the following provides a reliable fit to the relic density of 
the symmetric component {\it in the absence of an asymmetric component}:
\be
(\Omega h^2)_{\chi+\bar\chi} \simeq  0.19 \left(\frac{m_\psi/\lambda_3^2}{500\,\mathrm{GeV}}\right)^{1.82}
\label{eq:symm}
\ee
The additional presence of the asymmetric component, however, makes
the annihilation of $\bar\chi$ more efficient. From Fig.~4 in Ref.~\cite{Graesser:2011wi},
we can read off that already an annihilation rate $\langle\sigma v\rangle$ a factor 
of 2 larger than the cross section $\langle\sigma v\rangle_0$ that would give the
correct relic density in the purely symmetric case, 
i.e.~in accordance with Eq.~(\ref{eq:symm}), results in 
a suppression of the symmetric component by a factor of almost 50. 
Choosing this reference relic density as $(\Omega h^2)_{\chi+\bar\chi}=0.145$
(rather than the standard value of $(\Omega h^2)_{\rm DM}\simeq0.12$ 
because 17\,\% of the DM will convert to neutrons), this translates to the requirement
\be
	{m_\psi\over\lambda_3^2} \lesssim 300\,{\rm GeV}\,.
\label{psibound}
\ee
This condition is easily satisfied as long as $\lambda_3$ is not
too small (note that decreasing $m_\psi$ below about 100\,GeV 
would increases the tension 
with $\Delta N_\mathrm{eff}$, c.f.~Fig.~\ref{fig:Neff}).

Even after freeze-out, both the symmetric and the asymmetric DM
component are kept in local (kinetic) equilibrium with $\nu'$. 
This is exactly one 
of the scenarios studied in Ref.~\cite{Bringmann:2016ilk}
where, due to the presence of a light mediator, elastic scattering between 
$\chi$ and $\nu'$ is highly efficient and  kinetic decoupling therefore 
happens significantly
later than for standard WIMP candidates. The dark acoustic 
oscillations in the $\chi$-$\nu'$ fluid during and after decoupling lead 
to a cutoff in the power spectrum of matter density perturbations 
with an associated mass scale of
\bea
 M_\mathrm{cut} &\simeq& 2\cdot10^9 \,M_\odot\, \left(\frac{\xi}{0.5}\right)^\frac92 Q_{\nu'}^\frac32 \left(\frac{g'}{0.001}\right)^3 \left(\frac{m_{\gamma'}}{60\,\mathrm{keV}}\right)^{-3}.\nonumber\\
\label{eq:mcut}
\eea
A warm DM candidate with mass
$\left[10^{11}h^{-1}M_\odot/M_\mathrm{cut}
\right]^{1/4}$ keV  generates an almost identical small-scale
suppression of the nonlinear power  spectrum
\cite{Vogelsberger:2015gpr}. This implies that Lyman-$\alpha$
constraints on  our model can be easily evaded as long as the charge
of the dark sector neutrino is not  too large,  $Q_{\nu'}\lesssim1$
\cite{Irsic:2017ixq,Garny:2018byk,Garzilli:2018jqh}.  For
$Q_{\nu'}\sim1$,   in fact, the resulting mild suppression of the
power spectrum might  help to alleviate the missing satellites problem
\cite{Bringmann:2016ilk,Huo:2017vef}.

\begin{figure}
\centerline{\includegraphics[width=0.45\textwidth]{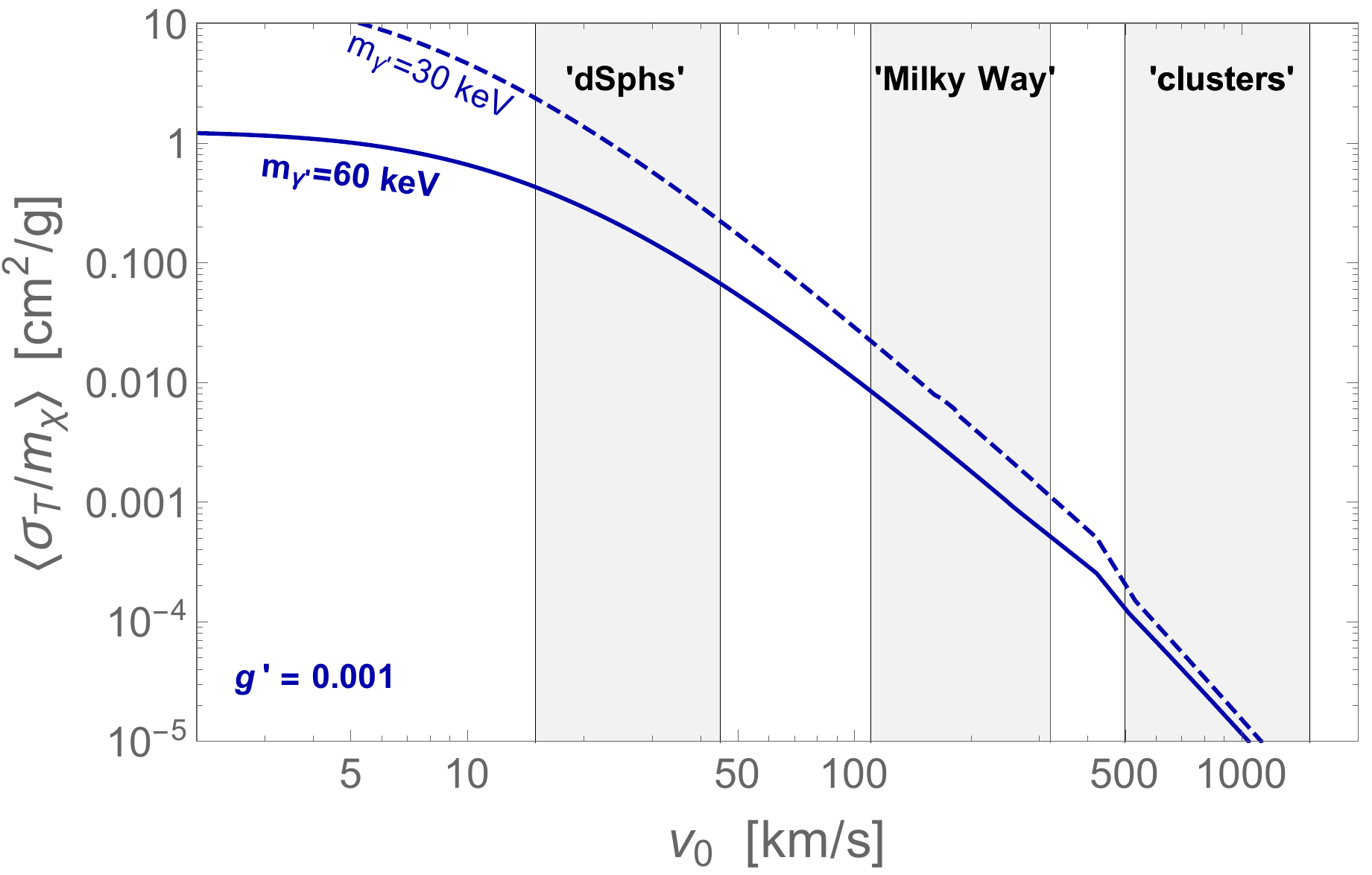}}
\caption{Phase space averaged self-interaction cross section over
mass, $\langle\sigma_T\rangle/m_\chi$, versus characteristic
DM velocity $v_0$, for $g'=0.001$ and $m_{\gamma'}=60$\,keV
(solid) or $m_{\gamma'}=30$\,keV (dashed).
The vertical bands indicate typical velocities
encountered in dwarf spheroidal galaxies, Milky Way-sized galaxies,
and clusters. Mitigating the $\Lambda$CDM small-scale problems
requires $\langle\sigma_T\rangle/m_\chi\sim 0.1-1$\,cm$^2$/g
at dwarf scales.}
\label{fig:sigma}
\end{figure}  

\subsection{Dark matter self interactions}
\label{sec:sub}
The cross section for $\chi\chi\to\chi\chi$ scattering is nonperturbatively
enhanced by multiple $\gamma'$ exchange, and hence cannot be 
calculated in the Born limit.  We use \ds\ to calculate 
the transfer cross section for a repulsive Yukawa potential 
in the classical limit $m_\chi v_\mathrm{rel}\gg m_{\gamma'}$,
based on parameterizations from Ref.~\cite{Cyr-Racine:2015ihg},
averaged over a Maxwellian velocity distribution with a most probable 
velocity $v_0$. 
For our benchmark model 
(\ref{benchmark}) 
and $v_0=30$\,km/s, this leads to $\langle\sigma_T\rangle 
/ m_\chi\sim 0.15$\,cm$^2$/g,
which may be strong enough to visibly affect the structure of 
dwarf galaxies \cite{Spergel:1999mh,Tulin:2017ara,Karananas:2018goc}.
The cross sections required to fully address the small-scale 
structure problems of cold DM are typically quoted to be a bit larger. 
This can be achieved by choosing a slightly smaller dark photon
mass: for $m_{\gamma'}=30$\,keV for example, still respecting
the constraint on $M_{\rm cut}$ implied by Eq.~(\ref{eq:mcut}),
we find $\langle\sigma_T\rangle / m_\chi\sim 0.6$\,cm$^2$/g.
The dependence of the transfer cross section
on $v_0$ is shown in Fig.~\ref{fig:sigma}. The self-interaction
rate drops very sharply with the typical velocity, implying that 
constraints on cluster 
scales~\cite{Randall:2007ph,Harvey:2015hha,Kaplinghat:2015aga} 
are not relevant for our model.
\bigskip

\section{Conclusions}

We have demonstrated that baryogenesis, starting from an
initial dark matter asymmetry, can be successfully implemented 
through dark matter-neutron oscillations.  It is intriguing that
the same model invented to address the neutron lifetime anomaly
in Ref.\ \cite{Cline:2018ami} is capable, with small modifications,
of producing the observed baryon asymmetry in this way.  A key
requirement is that a somewhat smaller mass splitting $\Delta m =
m_n-m_\chi\sim 0.1\,$MeV is now needed, relative to resolving only
the lifetime anomaly.  This arises because the baryogenesis curves
in Fig.~\ref{fig:cont} fall in the region excluded by $n\to\chi\gamma$
limits at larger $\Delta m$.  Simultaneously satisfying the neutron
lifetime anomaly requires small $g'\sim 10^{-3}$ and a 
consequently lower dark photon mass,
$m_\gamma' \sim 60\,$keV, mandating an additional dark radiation species
$\nu'$ so that dark photons can decay fast enough to avoid problems
with BBN and the CMB.  Intriguingly, the combination of dark radiation
and a very light dark photon may help to alleviate known
issues of $\Lambda$CDM cosmology at small scales.

The most ambitious version of our scenario, where both baryogenesis
and the neutron lifetime anomaly are treated, is challenged by extra
radiation contributing to $N_{\rm eff}$, being near the edge of
current  cosmological constraints. Updated BBN constraints and future
CMB observations, like the planned Simons
observatory~\cite{Ade:2018sbj},  are likely to provide the most
sensitive experimental test of the  proposal.  On the other hand, no dark
radiation is needed in a simpler version,  
exemplified by the parameter choices (\ref{benchmark2}), that provides only
baryogenesis but no significant dark decay  channel for the neutron. 
In this version, new physics signals could arise from kinetic mixing
of the photon with the dark photon.  In both cases it is possible that
the heavy scalar triplets required by the UV complete version of the
model may be accessible at the LHC.

Let us stress in closing that the main phenomenon, of
asymmetry-sharing through $n$-$\chi$ oscillations, depends only upon 
the existence of $n$-$\chi$ mass mixing at low energies.  It is
remarkable that purely standard model physics, giving the neutron a 
negative self-energy correction at finite temperature, makes it
possible for the oscillations to be resonant at temperatures not far
above BBN.  To make this part of a coherent picture including dark
matter generation in the early universe requires lifting the effective
theory to its UV completion.   It could be interesting to find other
examples of microscopic models leading to these phenomena.

\acknowledgements
We thank  F.\ Kahlhoefer and K.\ Kainulainen for valuable discussions.
This work was performed in part at the Aspen Center for Physics, which is supported by National Science Foundation grant PHY-1607611.
T.B.\ wishes to thank McGill University, where part of this manuscript was 
completed, for generous support and hospitality.
J.M.C.\ and J.M.C.\ gratefully acknowledge funding from the Natural Sciences and Engineering Research Council of Canada (NSERC).


\begin{appendix}

\section{Thermal self-energy of $\chi$}
\label{appA}

\begin{figure}[t]
\centerline{\includegraphics[width=0.45\textwidth]{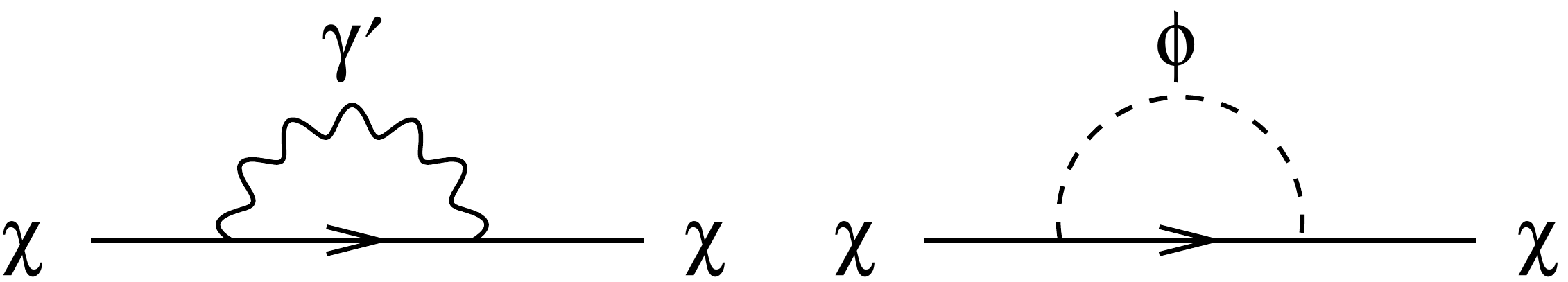}}
\caption{Thermal self-energy corrections to $\chi$, where the
intermediate boson has a finite-temperature propagator.}
\label{fig:loop}
\end{figure}  

The thermal contributions to the $\chi$ self-energy are shown in 
Fig.~\ref{fig:loop}, in which the propagators of the intermediate
light bosons are replaced by their thermal corrections (see for
example
ref.\ \cite{Notzold:1987ik}),
\be
	{1\over p^2 - m^2} \to {2\pi \delta(p^2-m^2)\over
				e^{E/T}-1}\,.
\ee
Evaluating the contribution from the 
hidden photon $\gamma'$ gives
\be
	 \Sigma_\chi 
	= g^2\int {d^{\,4}p\over (2\pi)^3}\,
	\gamma^\mu {\delta(p^2 - m^2_{\gamma'})\over \slashed{k} + \slashed{p} - m_\chi}
	\gamma^\nu {-\eta_{\mu\nu} + p_\mu p_\nu/m^2_{\gamma'}\over
	e^{|p_0|/T_{\gamma'}}-1}\,,
\ee
which simplifies in the limit $\vec k\to
0$ of nonrelativistic $\chi$, and neglecting the dark photon 
mass, $m_{\gamma'}\ll T_{\gamma'}$.  We get
\be
	\Delta E_\chi=\Sigma_\chi\cong  {g'}^2 {T_{\gamma'}^2\over 8 m_\chi}
\label{eq:sigmachi}	
\ee
in this limit.

A similar contribution to $\Delta E_\chi\sim \lambda_3^2/8m_\Psi$ arises when 
introducing the additional fermion $\psi$, with a coupling to $\chi$ given in 
Eq.~(\ref{psi_int}). For large values of $g'$, this contribution is parametrically 
suppressed by a factor of $m_\chi/m_\psi$ with respect to Eq.~(\ref{eq:sigmachi}). 
If $g'\ll  \lambda_3$, on the other hand, this can in fact be the dominant 
contribution to the thermal self-energy of $\chi$ -- but would still be so small 
that it does not affect any of our numerical results (in Fig.~\ref{fig:NSE}, in particular, 
we would still remain on the curve labelled `$\xi g'=0$').
In our analysis, we therefore neglect the contribution from $\psi$ to $\Delta E_\chi$.

\section{Decoupling of the dark sector}
\label{dim7freezeout}
We estimate the freezeout temperature for the dimension-7 interaction
(\ref{lowEint}) from the scattering process $\phi\chi\to udd$.  At
$T\gg m_\chi$, we can approximate all particles as being massless, with momenta
$p_1 = p_\chi$, $p_2 = p_\phi$, and $p_{3,4,5}$ for the final-state
quarks.  The matrix element is
\[
	\langle |{\cal M}|^2\rangle = {6\over\Lambda^6}\,p_1\cdot p_3\,
	p_4\cdot p_5
\]
where $1/\Lambda^3\cong 1.7\times 10^{-10}\,{\rm GeV}^{-3}$ is the coefficient of the operator
(\ref{lowEint}).  We can evaluate the cross section in the center of
mass frame, choosing the 3-momenta to be 
\bea
	\vec p_1 &=& E_1(s_\alpha c_\phi, s_\alpha s_\phi, c_\alpha)\\
	\vec p_2 &=& -\vec p_1\\
	\vec p_3 &=& E_3(0,0,1)\\
	\vec p_4 &=& E_4(s_\gamma,0,c_\gamma)\\
	\vec p_5 &=& -(\vec p_3 + \vec p_4)
\eea
where the angle between $p_3$ and $p_4$ is fixed by energy-momentum
conservation,
\be
	c_\gamma =  1 + {s - 2\sqrt{s}(E_3+E_4)\over 2 E_3 E_4}
\ee
and $E_5 = \sqrt{s} - E_3 - E_4$.   Then $p_1\cdot p_3\,p_4\cdot p_5
= s(1-c_\alpha)E_3(\sqrt{s}-2 E_3)/4$, and  
the cross section is given by
\bea
	\sigma &=& {1\over 16 (2\pi)^4\,s}\int d c_\alpha d\phi
	\int_0^{\sqrt{s}/2} dE_3\int_{\sqrt{s}/2-E_3}^{\sqrt{s}/2}
	dE_4\, \langle |{\cal M}|^2\rangle \nn\\
	&=& {s^2\over 1024\,\pi^3\Lambda^6}
	\label{eq:sigdec}
\eea
The thermal average is
\be
	\langle\sigma v\rangle = {1\over 32\, T^5}\int_0^\infty ds\,
	\sigma s^{3/2} K_1(\sqrt{s}/T) = 
	{9\, T^4\over 256\pi^3\,\Lambda^6} 
\ee
Multiplying by the thermal density of a complex scalar and equating the
scattering rate to the Hubble rate yields the freezeout temperature
$T_f = 11\,$GeV.

\section{Relations between chemical potentials and initial asymmetries}
\label{app:ChemPot}

In this appendix we present a calculation of the chemical potentials of both standard model and dark sector species when there is no baryon number violating interaction with $\psi'$ which sets $\mu_\chi = \mu_\phi$. After the electroweak phase transition, the chemical potentials of the particle species in equilibrium are given by the equations in Sec.~\ref{sec:seq}, with an additional relation derived from the fact the difference between total baryon number and lepton number is conserved when $\psi'$ is not present: 
\be
	\label{eq:muBminusL}
	2 \mu_\chi + B - L = 2 \mu_0 \, .
\ee
Here $B = 6 (\mu_u + \mu_d)$ and $L = \mu_\nu + 2 \mu_\ell$ where $\mu_\ell$ is the sum of the charged lepton chemical potentials. $\mu_0$ is the initial $\chi$ chemical potential. 

\newpage
From these equations, the chemical potentials of the various species can be determined in terms of the initial asymmetries:
\begin{subequations}
\label{eq:ChemPot}
\bea
	B &=& \frac{18}{61} (2 \mu_0 + Q'_{\nu'} \mu_{\nu'}) \\
	L &=& -\frac{25}{12} B \qquad 6 \mu_W = \frac{2}{3} B \\
	2 \mu_\chi &=& \frac{1}{122} (22 \mu_0 - 111 Q'_{\nu'} \mu_{\nu'}) \\
	2 \mu_\phi &=& \frac{1}{122} (-22 \mu_0 - 133 Q'_{\nu'} \mu_{\nu'})
\eea
\end{subequations}
The coefficients of the chemical potentials in the above expressions are chosen so that their ratios correspond to the ratios of the asymmetries, \textit{e.g.} $2 \mu_\chi / B = (n_\chi - n_{\bar \chi}) / (n_B - n_{\bar B})$. If $Q'_{\nu'} \mu_{\nu'} = -2 \mu_0$, then $B = L = \mu_W = 0$, $2 \mu_\chi = 2 \mu_\phi = 2 \mu_0$, and the analysis of Sec.\ \ref{sec:osc} directly applies. However, other values for the initial asymmetries can lead to a baryon asymmetry when these interactions freeze out that is greater than what is observed today. Such scenarios are not viable -- the subsequent oscillations will only drive the ratio of baryon to $\chi$ asymmetries to 1, rather than reducing the baryon asymmetry to the observed value.

\section{Limit on the $\nu'$ mass}
\label{sec:numass}

Generically, one needs to introduce dark sector particles $\nu'$ (though not necessarily massless fermions) 
with a non-vanishing chemical potential, 
which are stable because they carry a charge $Q'\neq1$. This extra field is needed to ensure that the universe is neutral in U(1)$'$ charge at early times. These new states $\nu'$ should not be too heavy in order not to overclose 
the universe.
For the case of benchmark scenario \textbf{B2}, we can estimate this constraint by noting
that the annihilation into $\gamma'\gamma'$ keeps both $\nu'$ and $\chi$ in
equilibrium, respectively, until they are non-relativistic. This allows to relate 
$\mu_\nu$ to $n_\nu-n_{\bar\nu}$ in the non-relativistic limit until freeze-out at 
$T_{\nu, \mathrm{fo}}$, after which the comoving number density stays constant.
The same applies, correspondingly, to the $\chi$ particles. For chemical potentials of 
roughly the same size,  $\left|\mu_\chi/\mu_\nu\right|\sim\mathcal{O}(1)$, we
then find that $m_{\nu'}\sim100$\,MeV is already sufficient to suppress
$\rho_\nu/\rho_\chi\ll 1$ at late times. While the existence of the additional states 
$\nu'$ is necessary for our baryogenesis scenario to work, they can thus 
be heavy enough to leave the phenomenology of \textbf{B2} unaffected.

\end{appendix}

\end{document}